\documentclass[aps,prc,preprint,showpacs,showkeys]{revtex4}
\usepackage{graphicx}
\usepackage{bm}
\usepackage{epsf}

\begin{document}

\title{Medium modifications of baryon properties in nuclear matter and hypernuclei}
\author{J. S. Liang}
\affiliation{School of Physics, Nankai University, Tianjin 300071, China}
\author{H. Shen}
\email{shennankai@gmail.com}
\affiliation{School of Physics, Nankai University, Tianjin 300071, China}

\begin{abstract}
We study the medium modifications of baryon properties in nuclear many-body
systems, especially in $\Lambda$ hypernuclei.
The nucleon and the $\Lambda$ hyperon are described in the Friedberg-Lee model
as nontopological solitons which interact through the self-consistent exchange
of scalar and vector mesons. The quark degrees of freedom are explicitly
considered in the model, so that the medium effects on baryons could be
investigated. It is found that the model can provide reasonable descriptions
for nuclear matter, finite nuclei, and $\Lambda$ hypernuclei.
The present model predicts a significant increase of the baryon radius
in nuclear medium.
\end{abstract}

\pacs{12.39.-x, 21.60.-n, 21.80.+a, 24.10.Jv }
\keywords{Friedberg-Lee model, Nuclear matter, Medium Effect, Hypernuclei}
\maketitle



\section{Introduction}
\label{sec:1}

The change of hadron properties in nuclear medium
is a very active field of experimental and theoretical research.
There are many experimental evidences indicating that the
properties of the nucleon bound in a nucleus
differ from those of a free nucleon.
The famous European Muon Collaboration (EMC) effect implies that
the nucleon structure function in nuclei deviate from those in
a free nucleon~\cite{emc1}.
Recent experiments at Jefferson Laboratory have provided precise
measurements of the EMC effect in light nuclei~\cite{emc2,emc3}.
On the other hand, there are many theoretical works on the study of
in-medium hadron properties based on various models~\cite{prc85,qmf98,npa00,ppnp07}.
The quark-meson coupling (QMC) model proposed by Guichon~\cite{qmc88}
opens the possibility of understanding the change of the nucleon internal structure
in nuclei based on quark degrees of freedom.
The QMC model is considered as an extension of the successful
treatment of nuclear many-body systems at the hadron level,
known as the relativistic mean-field (RMF) model~\cite{WS86}.
The QMC model describes the nuclear system as nonoverlapping MIT bags
in which the confined quarks interact through the self-consistent exchange
of scalar and vector mesons in the mean-field approximation.
In the QMC model, the quark structure of the nucleon plays a crucial role
in the description of nuclear matter and finite nuclei~\cite{ppnp07}.
The QMC model could be used to study the medium modifications of
nucleons and mesons~\cite{qmc95,qmc98}.
In the past few decades, the QMC model has been extensively
developed and applied to various nuclear
phenomena~\cite{ppnp07,qmc96,qmccm1,qmc97,qmcl1,qmcl2,qmc99,qmc04,shen08}.
There are other models that incorporate quark degrees of freedom in
the study of nuclear many-body systems.
The quark mean-field (QMF) model~\cite{qmf98} uses the constituent quark model
for the nucleon instead of the MIT bag model,
where the constituent quarks interact with the meson fields created
by other nucleons. The QMF model has been successfully used for the description
of nuclear matter, finite nuclei, and hypernuclei~\cite{qmf00,qmf02,qmf05}.
Recently, the QMF model has been extended to a model based on
$\text{SU(3)}_{L}\times \text{SU(3)}_{R}$ symmetry and scale invariance~\cite{qmf0205}.
Using the Nambu-Jona-Lasinio model to describe the nucleon as a
quark-diquark state, it is also possible to discuss the stability of nuclear
matter based on the QMC idea~\cite{njl01}.
The advantage of these models is their simplicity and self-consistency
to incorporate quark degrees of freedom in the study of nuclear
many-body systems.

In this paper, we prefer to use the Friedberg-Lee (FL) model~\cite{fl77}
to describe the baryons in nuclear matter and $\Lambda$ hypernuclei.
In the FL model, the baryon is described as a bound state of three quarks
in a nontopological soliton formed by a scalar composite gluon field.
The FL model has the advantages that it is manifestly covariant and
it exhibits dynamical bag formation due to the coupling of quarks to
the phenomenological scalar field. Furthermore, it includes the MIT
bag as a special case. The FL model has the structure of a simple
field theory with quarks coupled to the scalar field, and the source
of the scalar field is the quark scalar density. Therefore, the
coupled equations of this model must be solved in a self-consistent
manner. The structure of the baryon is fully dynamical in the FL model,
so its response to external fields can be calculated consistently.
The attractive feature of the FL model makes it suitable for the study
of the medium modifications of baryon properties. Also, it could
be used to investigate the deconfinement phase transition at
high densities and/or temperatures based on its dynamical formation
of solitons. In past decades, the FL model has been extensively
used to study the properties of hadrons in free
space~\cite{fl82,fl83,Lubeck86,Lubeck87,fl86,fl90w,fl90,fl92,su05}.
It has also been applied to study the medium modifications of the
nucleon~\cite{prc85,jpg93} and dense-matter properties~\cite{npa00,fl10}.
In our previous work~\cite{wen08}, we studied the properties of nuclear matter
and finite nuclei by describing the nuclear many-body system
as a collection of nontopological solitons within the FL model.
The quarks inside the soliton bag couple not only to the
scalar composite gluon field that binds the quarks together into nucleons
but also to additional meson fields generated by the nuclear environment.
The nucleons interact through the self-consistent exchange of these mesons,
which are treated as classical fields in the spirit of the QMC and QMF models.
The properties of finite nuclei and inner nucleons have been investigated
in Ref.~\cite{wen08}. The main purpose of the present work is to extend
the model to flavor SU(3) and further to the study of $\Lambda$ hypernuclei.
Considering the importance and complexity of medium modifications of baryons,
it is interesting to study the change of baryon properties in nuclear medium
based on various feasible methods and compare their results.

Extensive efforts have been devoted to the study of hypernuclei,
which play an important role in strangeness nuclear physics.
Among many strange nuclear systems, the single-$\Lambda$ hypernucleus
is the most investigated one~\cite{lexp06}. There exist many experimental data
for various single-$\Lambda$ hypernuclei over almost the whole mass
table~\cite{lexp06,lexp88,lexp91,lexp96}, while there are few experimental data
concerning other hypernuclei. Theoretically, the properties of
$\Lambda$ hypernuclei have been investigated based on various models.
The RMF models have been successfully applied to describe hypernuclei
with adjustable meson-hyperon couplings and tensor
couplings~\cite{rmf2,rmf3,rmf4,rmf5,shen06}.
The properties of $\Lambda$ hypernuclei have also been examined
in the QMC and QMF models~\cite{qmcl1,qmcl2,qmf02},
where both the nucleon and $\Lambda$ are the composites of three quarks.
For $\Sigma$ hypernuclei, the situation is quite different.
So far, only one bound $\Sigma$ hypernucleus, $^{4}_\Sigma\rm{He}$,
was detected~\cite{PRL98}, despite extensive searches.
The analysis of $\Sigma$ atomic experimental data suggests
that $\Sigma$-nucleus potentials have a repulsion inside
the nuclear surface and an attraction outside the nucleus with
a sizable absorption. In the case of $\Xi$ hypernuclei,
although there is no definitive data for any $\Xi$ hypernucleus
at present, several experimental results suggest that
$\Xi$-nucleus interactions are weakly attractive~\cite{PRC00xi,Hiyama2010}.
It is a challenge to naturally explain the attraction in
$\Lambda$-nucleus potentials, and at the same time the repulsion
in $\Sigma$-nucleus potentials.
In general, the RMF models use some suggested potential values
to determine the meson-hyperon couplings~\cite{shen02,jpg08},
which play important roles in hypernuclear physics and neutron
star properties. The QMC model has been applied to a systematic
study of $\Lambda$, $\Sigma$, and $\Xi$ hypernuclei
in Ref.~\cite{qmcl2}, where $\Sigma$ hypernuclei could be
bound by an amount similar to $\Lambda$ hypernuclei.
Recently, the QMC model has been extended to include the
quark-quark hyperfine interactions due to gluon and pion
exchanges~\cite{qmc07,qmc08,cqmc08,cqmc09}. It was found that
the hyperfine interaction due to the gluon exchange plays an important
role in the in-medium baryon spectra, while the pion-cloud effect
is relatively small~\cite{cqmc09}. Furthermore, the hyperfine
interaction can make the equation of state hard and thus
enhance the mass of a neutron star~\cite{qmc07,cqmc12}.
The medium effect on the hyperfine interaction increases
the splitting between the $\Lambda$ and $\Sigma$ masses as the density
rises, and $\Sigma$ hypernuclei are found to be unbound in Ref.~\cite{qmc08},
which is completely different from the prediction of Ref.~\cite{qmcl2}.
In the present work, we focus on the description of $\Lambda$ hypernuclei
with the baryons described by the FL model, which is an extension of
our previous work~\cite{wen08}. In addition, a primary application of the
present model to $\Sigma$ and $\Xi$ hypernuclei is briefly discussed.

The outline of this paper is as follows. In Sec.~\ref{sec:2}
we briefly describe the FL model both in free space and in nuclear medium.
In Sec.~\ref{sec:3}, we show the calculated results of $\Lambda$ hypernuclei
and the medium modifications of the nucleon and $\Lambda$ in nuclear matter
and hypernuclei. Section~\ref{sec:4} is devoted to a summary.

\section{Formalism}
\label{sec:2}

In this section, we first give a brief description of the FL
model. Then we describe nuclear matter as a collection of
nontopological solitons and study the properties of nucleons and $\Lambda$
hyperons inside nuclear matter. The model can be applied to study
finite systems such as nuclei and $\Lambda$ hypernuclei.

\subsection{Baryons in the FL model}
\label{sec:2a}

The nucleon and the $\Lambda$ hyperon as composites of three quarks
are described in terms of the FL model, in which three quarks
couple with a phenomenological scalar field to form a nontopological soliton bag.
The effective Lagrangian of the FL model is written as
\begin{equation}
\mathcal{L}=\sum_{f=u,d,s}
 \bar{\psi}_f\left( i\gamma_{\mu}\partial^{\mu}-m_f-g_\phi^f \phi\right)\psi_f
+\frac{1}{2}\partial_{\mu }\phi \partial ^{\mu }\phi -U\left({\phi}\right) ,
\label{eq:lag0}
\end{equation}
where $\psi_f$ denotes the quark field of flavor $f$.
We take the quark masses $m_u=m_d=5$ MeV  and $m_s=190$ MeV.
$\phi$ is a color-singlet scalar field
that can be interpreted as the phenomenological representation of quantum
excitations of the self-interacting gluon field. The self-interaction of the
scalar field is described by the potential
\begin{equation}
U(\phi )=\frac{a}{2!}{\phi }^{2}+\frac{b}{3!}{\phi }^{3}+\frac{c}{4!}{\phi }%
^{4}+P ,  \label{eq:pot}
\end{equation}%
where the parameters $a$, $b$, and $c$ are within a range so that $U({\phi })$
has a local minimum at $\phi =0$ and a global minimum at $\phi =\phi_{v}$.
The constant $P$ is determined to make $U({\phi_{v}})=0$, and then the value
$U(0)=P$ is to be identified with the volume energy density of the bag.
$\phi_{v}$ is the value of the soliton field in the physical vacuum
where the quark of flavor $f$ has the effective mass $m_f+g_\phi^f \phi_{v}$.
In the perturbative vacuum where valence quarks exist,
the soliton field $\phi$ is reduced to be near zero.

The quark field operator is expanded in a complete orthogonal set of
Dirac spinor functions as $\psi_f=\sum\limits_{i} b_i^f \psi_i^f $,
where $b_i^f$ is the annihilation operator of the quark with flavor $f=u,d,s$.
The soliton field is treated as a classical field
that is a time-independent \emph{c}-number field $\phi \left(\mathbf{r}\right)$.
In the nucleon or $\Lambda$, the valence quarks are in the lowest Dirac
state $\psi_0^f$, then $\phi$ and $\psi_0^f\; (f=u,d,s)$ satisfy
the coupled differential equations
\begin{equation}
\left[-i\mathbf{\alpha}\cdot \mathbf{\nabla}
      +\beta\left(m_f+g_\phi^f \phi\right)\right]\psi_0^f
=\epsilon_{0}^f\psi_0^f ,  \label{eq:free1}
\end{equation}
\begin{equation}
-\nabla^{2}\phi +\frac{\partial U(\phi)}{\partial \phi}
=-\sum_{f} g_\phi^f \bar{\psi}_{0}^f\psi_{0}^f.  \label{eq:free2}
\end{equation}
The coupled equations have to be solved numerically.
The lowest Dirac state has the form
\begin{equation}
\psi_{0}^f=\left(
\begin{array}{c}
u\left( r\right) \\
i\mathbf{\sigma}\cdot \hat{\mathbf{r}} v\left( r\right)%
\end{array}%
\right) \chi ,  \label{eq:wf}
\end{equation}%
with $\chi$ being the Pauli spinor. The total energy of the baryon $B$ ($B=N,\,\Lambda$)
is given by
\begin{equation}
E_B=\sum_{f}\epsilon_{0}^f+4\pi \int dr \ r^{2}\left[ \frac{1}{2}\left( \frac{d\phi }
  {dr}\right) ^{2}+U(\phi )\right] .  \label{eq:ne}
\end{equation}

There are several methods for the removal of the center-of-mass motion in the FL
model~\cite{fl83,Lubeck86,Lubeck87}. In our previous work~\cite{wen08},
we used two approaches for the center-of-mass correction.
The first one is based on the relativistic energy-momentum relation~\cite{fl83}
and the second one is the Peierls-Yoccoz projection technique~\cite{Lubeck86}.
For simplicity, in this work we take only the first approach for the center-of-mass
correction that has been extensively discussed in Refs.~\cite{fl83,wen08}.
The rest mass of the baryon is given by
\begin{equation}
M_B=\sqrt{E_B^{2}-\sum_{f}\langle \mathbf{p}_f^{2}\rangle }.  \label{eq:nm1}
\end{equation}
The corrected root-mean-squared (rms) radius is given by
\begin{equation}
r_{B}=\sqrt{\frac{1}{3}\sum_{f}\left[
1-\frac{2\epsilon_{0}^f}{E_B}+\frac{3{\epsilon_{0}^f}^{2}}{E_B^{2}}\right]
\langle r_f^{2}\rangle
+\frac{3}{2E_B^{2}}}.  \label{eq:nrc}
\end{equation}

In the present work, we take two sets of parameters in the FL
model which are constrained to reproduce the nucleon rms radius $r_{N}=0.83$ fm,
the nucleon mass $M_N=939$ MeV, and the $\Lambda$ mass $M_\Lambda=1115.7$ MeV.
Set A ($a=0$, $b=-8.71$ fm$^{-1}$, $c=57.76$, $g_\phi^{u}=g_\phi^{d}=16.7$,
$g_\phi^{s}=4.675$) is characterized by $a=0$,
where $U({\phi })$ has a inflection point at $\phi =0$.
Set B ($a=17$ fm$^{-2}$, $b=-289.048$ fm$^{-1}$, $c=1638.2$,
$g_\phi^{u}=g_\phi^{d}=20.345$, $g_\phi^{s}=5.52$)
is characterized by $P=0$.
We note that sets A and B correspond to the two limiting cases in the parameter space of
the FL model.

\subsection{Baryons in nuclear matter}
\label{sec:2b}

We describe nuclear matter as a collection of nontopological solitons.
The baryons interact through the self-consistent exchange of $\sigma$, $\omega$,
and $\rho$ mesons that are treated as classical fields in the mean-field
approximation. The quarks inside the baryon couple not only to the soliton
field $\phi$ which binds the quarks together
but also to additional meson fields $\sigma$, $\omega$, and $\rho$
generated by other baryons in nuclear matter.
We assume that the meson mean fields $\sigma$, $\omega$, and $\rho $ can be
regarded as constants in uniform matter and that the soliton field $\phi$
that serves to bind the quarks together does not participate in baryon-baryon
interactions. Therefore, $\phi$ depends on spatial coordinates inside the soliton
bag, whereas $\sigma $, $\omega $, and $\rho $ are constants. In nuclear matter,
the coupled Eqs.~(\ref{eq:free1}) and (\ref{eq:free2}) are expressed as
\begin{equation}
\left[-i\mathbf{\alpha}\cdot \mathbf{\nabla}
      +\beta\left(m_f+g_\phi^f \phi +g_{\sigma}^{f}\sigma\right)
      +g_{\omega }^{f}\omega +g_{\rho }^{f}\tau_{3}\rho \right]\psi_0^f
=\widetilde\epsilon_{0}^f \psi_0^f ,  \label{eq:im1}
\end{equation}
\begin{equation}
-\nabla^{2}\phi +\frac{\partial U(\phi)}{\partial \phi}
=-\sum_{f} g_\phi^f \bar{\psi}_{0}^f\psi_{0}^f,  \label{eq:im2}
\end{equation}
where $g_{\sigma }^{f}$, $g_{\omega}^{f}$, and $g_{\rho}^{f}$ are the
coupling constants of the $\sigma$, $\omega$, and $\rho$ mesons with the
quark of flavor $f$, respectively.
According to the Okubo-Zweig-Iizuka (OZI) rule, the nonstrange mesons
couple exclusively to the $u$ and $d$ quarks, not to the $s$ quark.
Therefore, we take $g_{\sigma }^{s}=g_{\omega}^{s}=g_{\rho}^{s}=0$ in the present work.
The couplings of these mesons to the $u$ and $d$ quarks are determined by
fitting saturation properties of nuclear matter.
In uniform matter, the constant $\sigma$ field provides an additional scalar
potential to the $u$ and $d$ quarks, and as a result the solutions
of Eqs.~(\ref{eq:im1}) and (\ref{eq:im2}) are affected by the $\sigma$ field.
On the other hand, the $\omega$ and $\rho$ fields do not cause any changes of
$\psi_{0}^f$ and $\phi$ except to shift the energy level by a constant vector
potential, $\widetilde{\epsilon}_{0}^f\left( \sigma ,\omega ,\rho \right)=
\epsilon_{0}^f\left(\sigma\right)+g_{\omega}^{f}\omega+g_{\rho}^{f}\tau_{3}\rho$.

Analogously to the case of free baryons, we consider the center-of-mass correction
and calculate the properties of baryons in nuclear matter.
The effective baryon mass is given by
\begin{equation}
M_B^{\ast}\left(\sigma\right)
=\sqrt{E_B^{2}-\sum_{f}\langle \mathbf{p}_f^{2}\rangle },
\label{eq:nms1}
\end{equation}
where
\begin{equation}
E_B=\sum_{f}\epsilon_{0}^f\left(\sigma\right)
+4\pi \int dr \ r^{2}\left[ \frac{1}{2}\left( \frac{d\phi }
  {dr}\right) ^{2}+U(\phi )\right].
\label{eq:imne}
\end{equation}
Since the quark wave function $\psi_{0}^f$ and the soliton field
$\phi$ are altered by the $\sigma$ mean field in nuclear matter,
the calculated baryon properties are different from those in free space.
They can be expressed as a function of the $\sigma$ mean field,
for instance, the effective baryon mass is given by Eq.~(\ref{eq:nms1}).

We use a hybrid treatment for nuclear matter, in which the effective masses
and couplings are obtained at the quark level,
whereas the baryon Fermi motion is treated at the hadron level.
The description of nuclear matter in this model has been presented
in Ref.~\cite{wen08}. Here, we only study the $\Lambda$ hyperon immersed
in nuclear medium, which satisfies the following Dirac equation:
\begin{equation}
\left[ i\gamma_{\mu }\partial^{\mu } - M^{\ast}_{\Lambda}\left(\sigma \right)
      -g_{\omega}^{\Lambda}\gamma^{0}\omega \right] \psi_{\Lambda} = 0 .
\label{eq:dirac0}
\end{equation}
The influence of the $\sigma$ meson on the $\Lambda$ hyperon is contained
in $M^{\ast}_{\Lambda}$ and generates the scalar potential
$U_s^{\Lambda}=M^{\ast}_{\Lambda}-M_{\Lambda}$.
The $\omega$ meson couples to the $\Lambda$ hyperon with the
coupling constant $g_{\omega}^{\Lambda}=2g_{\omega}^q\,(q=u,d)$
and generates the vector potential
$U_v^{\Lambda}=g_{\omega}^{\Lambda}\omega$.
In the present model, the mesons couple directly to the quarks
inside the baryons, and the quark-meson couplings are determined by
fitting saturation properties of nuclear matter.
Therefore, no more adjustable parameters exist when we study the
properties of the $\Lambda$ hyperon in nuclear matter.

\subsection{$\Lambda$ hypernuclei}
\label{sec:2c}

We treat a single $\Lambda$ hypernucleus as a system of many nucleons
and a $\Lambda$ hyperon which interact through the exchange of
$\sigma$, $\omega$, and $\rho$ mesons.
The nucleon and the $\Lambda$ hyperon as composites of three quarks
are described in terms of the FL model, and the mesons
couple directly to the quarks inside the baryons.
The effective baryon masses are obtained at the quark level,
whereas the baryon Fermi motion is treated at the hadron level.
To study the properties of $\Lambda$ hypernuclei, we start from
the effective Lagrangian at the hadron level within the mean-field approximation
\begin{eqnarray}
{\cal L} &=&
\bar\psi_N\left[ i\gamma_\mu\partial^\mu-M_N^{\ast}\left(\sigma\right)
-g_{\omega}^{N}\gamma^{0}\omega
-g_{\rho}^{N}\gamma^0\tau_3\rho
-e \gamma^0\frac{1+\tau_3}{2} A \right] \psi_N  \nonumber\\
 & &
+\bar\psi_{\Lambda} \left[ i\gamma_\mu\partial^\mu-M_{\Lambda}^{\ast}\left(\sigma\right)
-g_{\omega}^{\Lambda}\gamma^0\omega
\right] \psi_{\Lambda} \nonumber\\
 & &
-\frac{1}{2} (\bigtriangledown\sigma)^2
-\frac{1}{2} m_\sigma^2\sigma^2
+\frac{1}{2} (\bigtriangledown\omega)^2
+\frac{1}{2} m_\omega^2\omega^2
\nonumber\\
 & &
+\frac{1}{2} (\bigtriangledown\rho)^2
+\frac{1}{2} m_\rho^2\rho^2
+\frac{1}{2}(\bigtriangledown A)^2,
\end{eqnarray}
where $\psi_N$ and $\psi_{\Lambda}$ are the Dirac spinors for the
nucleon and $\Lambda$. The mean-field approximation
has been adopted for the exchanged $\sigma$, $\omega$, and $\rho$
mesons, while the mean-field values of these mesons are denoted by
$\sigma$, $\omega$, and $\rho$, respectively.  We take the meson
masses $m_{\sigma}=500$ MeV, $m_{\omega}=783$ MeV, and $m_{\rho}=770$ MeV.
Since the $\Lambda$ hyperon is neutral and isoscalar, it only couples
to the $\sigma$ and $\omega$ mesons. The influence of the $\sigma$ meson
is contained in $M^{\ast}_{N}$ and $M^{\ast}_{\Lambda}$.
For $\omega$ and $\rho$ mesons, the couplings at the hadron level are
related to those at the quark level by $g_{\omega}^{N}=3g_\omega^q$,
$g_{\omega}^{\Lambda}=2g_\omega^q$, and $g_{\rho}^N=g_\rho^q$, which
are based on a simple quark counting rule.

From the Lagrangian given above, we obtain the following
Euler-Lagrange equations:
\begin{eqnarray}
 & & \left[
i\gamma_{\mu}\partial^{\mu}-M_N^*
-g_{\omega}^{N}\gamma^{0}\omega
-g_{\rho}^{N}\gamma^0\tau_3\rho
-e \gamma^0\frac{1+\tau_3}{2} A \right] \psi_N
= 0,
\label{eq:A1}\\
 & & \left[
i\gamma_{\mu}\partial^{\mu}-M_\Lambda^*
-g_{\omega}^{\Lambda}\gamma^{0}\omega
\right]\psi_\Lambda
= 0,
\label{eq:A2}\\
 & &
 -\nabla^2\sigma + m_\sigma^2\sigma=
-\frac {\partial M_N^{\ast}}{\partial \sigma} \rho_s^N
-\frac {\partial M_\Lambda^*}{\partial \sigma} \rho_s^\Lambda,
\label{eq:A3}\\
 & &
 -\nabla^2\omega + m_\omega^2\omega=
g_\omega^N \rho_v^N +g^\Lambda_\omega \rho_v^\Lambda,
\label{eq:A4}\\
 & &
 -\nabla^2\rho + m_\rho^2\rho =
g_\rho^N \rho_3,
\label{eq:A5}\\
 & &
 -\nabla^2 A =
e \rho_p,
\label{eq:A6}
\end{eqnarray}
where $\rho_s^N$ ($\rho_s^\Lambda$), $\rho_v^N$ ($\rho_v^\Lambda$),
      $\rho_3$, and $\rho_p$
are the scalar, vector, third component of isovector, and proton
densities, respectively.
The mean-field values $\sigma$, $\omega$, $\rho $, and $A$
are functions of the spatial coordinates in a finite system, such as the $\Lambda$
hypernucleus. However, it is rather complicated to consider the
variation of these quantities over the small baryon volume.
Therefore, we take some suitably averaged form for the meson mean
fields in order to make the numerical solution feasible.
We use the local density approximation, which replaces the meson mean fields
by their value at the center of the baryon, and neglect the spatial variation
of the mean fields over the small baryon volume~\cite{qmc96,qmf00}.
Within this approximation, we solve the coupled Eqs.~(\ref{eq:A1})--(\ref{eq:A6})
for the $\Lambda$ hypernucleus,
where the effective masses $M_N^{*}$ and $M_{\Lambda}^{*}$
are obtained at the quark level.

\section{Results and discussion}
\label{sec:3}

In this section, we investigate the properties of nucleons and $\Lambda$
hyperons inside nuclear matter and $\Lambda$ hypernuclei.
The baryons as composites of three quarks are described in terms of
the FL model. We take two sets of parameters in the FL
model (sets A and B), which are constrained to reproduce the baryon
properties in free space as described in Sec.~\ref{sec:2a}.

To study the medium modifications of baryon properties, we take
the hybrid treatment, in which the effective baryon masses
are obtained at the quark level, whereas the baryon Fermi motion
is treated at the hadron level.
We describe nuclear matter as a collection of nontopological solitons.
The baryons interact through the self-consistent exchange of $\sigma$, $\omega$,
and $\rho$ mesons that are treated as classical fields in the mean-field
approximation. The mesons couple directly to the quarks
inside the baryons. The quark-meson couplings are determined by
fitting saturation properties of nuclear matter.
We list in Table~\ref{tab:matter} the quark-meson couplings and the nuclear
matter properties corresponding to the parameter sets A and B.
It is shown that the present model can provide a satisfactory description
of nuclear matter properties.
For the $\Lambda$ hyperon in nuclear matter, there is no more adjustable
parameters in the present model. We obtain the potential depth of a $\Lambda$
in saturated nuclear matter
to be around $-24$ MeV for set A and $-30$ MeV for set B.

\begin{figure}[htb]
\includegraphics[bb=30 90 515 470, width=8.6 cm, clip]{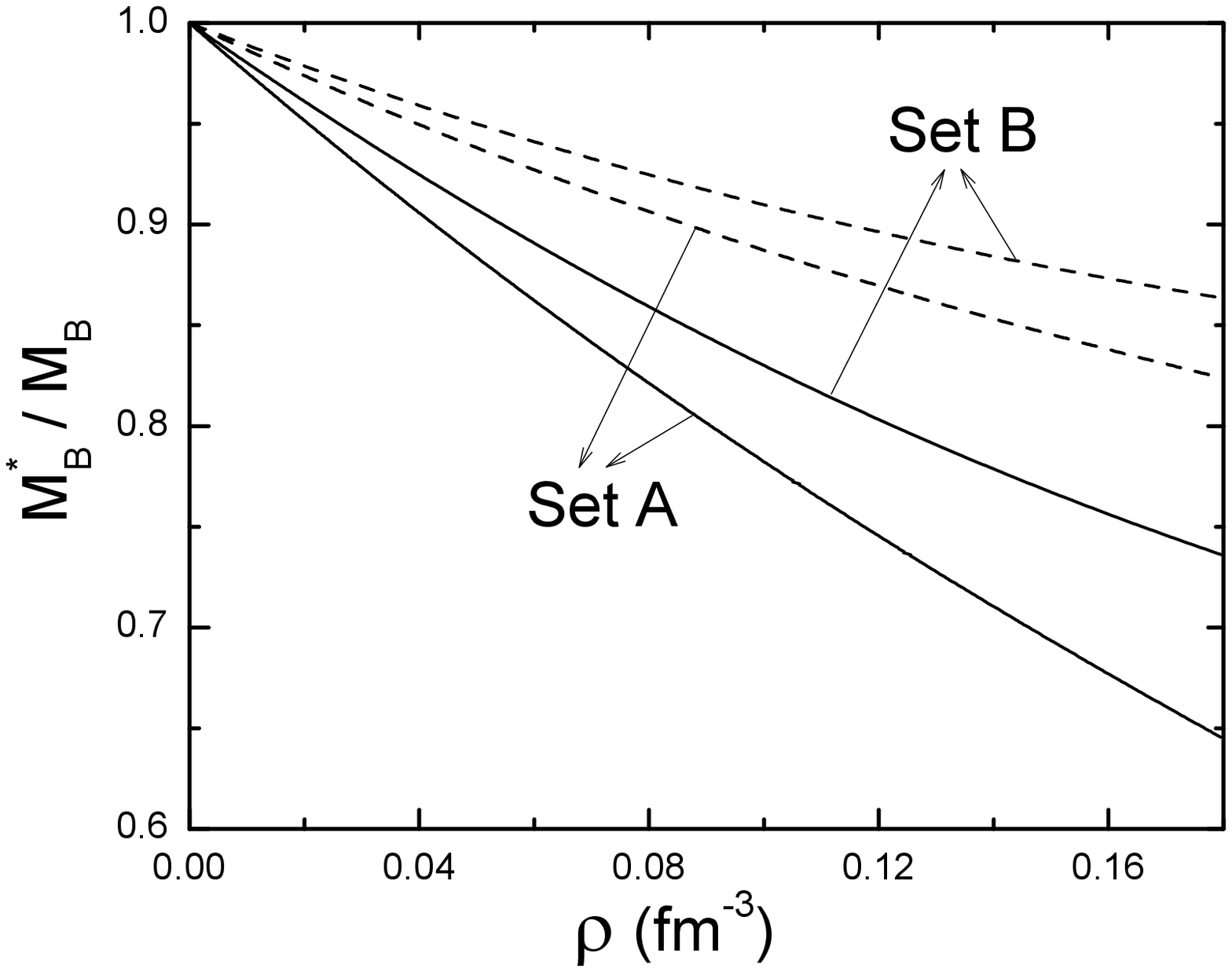}
\caption{The ratio of the effective baryon mass in nuclear matter
to that in free space $M_B^{\ast}/M_B$ ($B=N,\,\Lambda$) as a function
of the density $\rho$. The results of $N$ and $\Lambda$
are shown by the solid and dashed lines, respectively.}
\label{fig:mass}
\end{figure}

\begin{figure}[htb]
\includegraphics[bb=20 80 520 500, width=8.6 cm, clip]{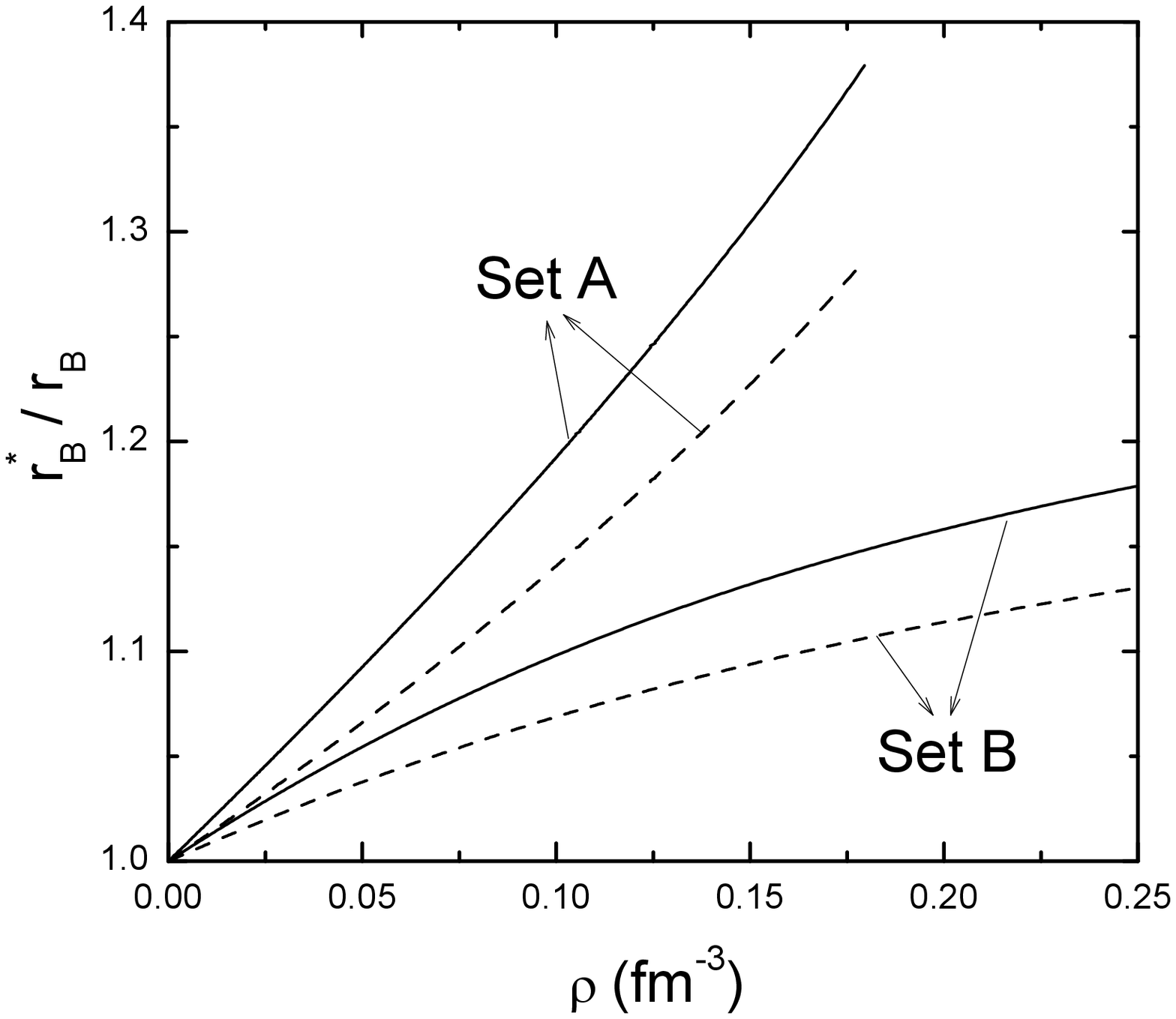}
\caption{The ratio of the baryon radius in nuclear matter
to that in free space $r_B^{\ast}/r_B$ ($B=N,\,\Lambda$)
as a function of the density $\rho$. The results of $N$ and $\Lambda$
are shown by the solid and dashed lines, respectively.}
\label{fig:r}
\end{figure}

In Fig.~\ref{fig:mass}, we present the ratio of the effective
baryon mass in nuclear matter to that in free space
$M_B^{\ast }/M_B$ ($B=N,\,\Lambda$) as a function of the
nuclear matter density $\rho$. The solid and dashed lines
correspond to the results of $N$ and $\Lambda$, respectively.
In the present model, the effective baryon mass is calculated
at the quark level, which is not a simple linear function of $\sigma$
as given in the RMF model~\cite{WS86}. It is more like the
characteristics of the QMC model~\cite{ppnp07}. As shown in Fig.~\ref{fig:mass},
both $M_N^*$ (solid lines) and $M_{\Lambda}^*$ (dashed lines)
decrease with increasing $\rho$.
It is seen that the reduction of $M_{\Lambda}^{*}$ is smaller
than that of $M_N^*$, because only two of the three quarks in
the $\Lambda$ hyperon are affected by the $\sigma$ mean field.
We note that the dependence of the effective baryon mass on the $\sigma$
mean field is calculated self-consistently within the FL model,
and therefore the ratio between the variation of the effective mass for
$\Lambda$ and that for $N$ is not as simple
as a constant obtained in the RMF models~\cite{rmf2,rmf3}.
In Fig.~\ref{fig:r}, we show the ratio of the baryon rms radius in
nuclear matter to that in free space $r_B^{\ast}/r_B$ ($B=N,\,\Lambda$)
as a function of the nuclear matter density $\rho$.
It is very interesting to see the expansion of the baryon size
in nuclear matter. The increase of the baryon size in the present model
is quite large. At normal nuclear matter density, we obtain
$r_N^{\ast}/r_N\sim 1.13 \; (1.30)$ and
$r_{\Lambda}^{\ast}/r_{\Lambda}\sim 1.09 \; (1.23)$ with the
parameter set B (A). The results of set B are similar to
that obtained in Ref. [37], while those of set A are apparently too large.
The main reason of the large expansion is due to the significant
modification of the quark wave function in nuclear medium,
which has been shown in Fig. 3 of Ref.~\cite{wen08}.
Compared with the nucleon, the $\Lambda$ hyperon
has less expansion since the wave function of the $s$ quark is not
obviously affected by the $\sigma$ mean field in nuclear matter.
The increase of the nucleon size in the present model is rather
different from those obtained in other models.
For example, the QMC model predicts only $1-3\%$
enhancement in the nucleon rms radius at normal nuclear matter
density~\cite{qmc96}, and the QMF model gives about $5-9\%$
increase~\cite{qmf00}. The chiral quark-soliton model predicts a
$2.4\%$ enhancement, while the swelling constrained by quasielastic
inclusive electron-nucleus scattering data is less than
$6\%$~\cite{cqs04}.

\begin{figure}[htb]
\includegraphics[bb=50 110 520 500, width=8.6 cm, clip]{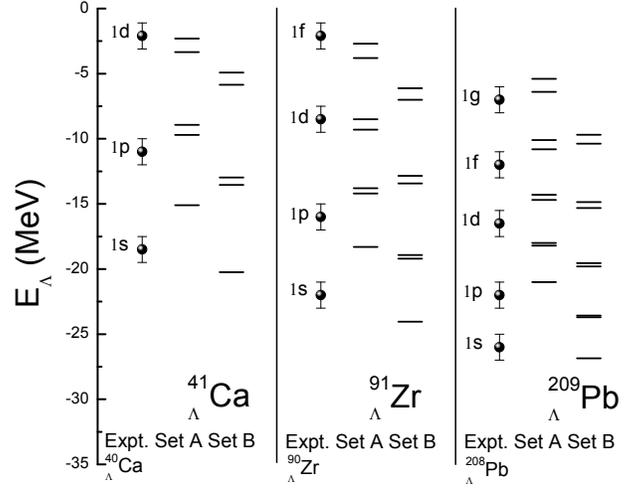}
\caption{The calculated $\Lambda$ single-particle energy $E_\Lambda$
    in $^{41}_\Lambda\rm{Ca}$, $^{91}_\Lambda\rm{Zr}$, and $^{209}_\Lambda\rm{Pb}$
    with the parameter sets A and B.
    The experimental data of $E_\Lambda$ in
    $^{40}_\Lambda\rm{Ca}$, $^{90}_\Lambda\rm{Zr}$, and $^{208}_\Lambda\rm{Pb}$
    are shown for comparison~\protect\cite{lexp91,lexp96}.}
\label{fig:spe}
\end{figure}

We study the properties of $\Lambda$ hypernuclei within the model
where the nucleon and $\Lambda$ are described in terms of
the FL model. In our previous work~\cite{wen08},
we studied the properties of finite nuclei and the modifications of
nucleon properties inside nuclei. Here we focus on the description of
$\Lambda$ hypernuclei. In Fig.~\ref{fig:spe}, we present the calculated
$\Lambda$ single-particle energy $E_\Lambda$ in several hypernuclei consisting
of a closed-shell nuclear core and a single $\Lambda$ hyperon.
The experimental values~\cite{lexp91,lexp96} are also shown for comparison.
It is found that the results of set B are closer to
the experimental values than those of set A. This is consistent with
the potential depth of the $\Lambda$ hyperon in nuclear matter,
which is found to be around $-24$ MeV for set A and $-30$ MeV for set B
at the saturation density. Generally, a $\Lambda$ potential around $-30$ MeV
is often used to constrain the meson-hyperon couplings in the RMF
model~\cite{PRC00}.
We note that there is no adjustable
parameter in the present calculation of $\Lambda$ hypernuclei.
This is different from the treatment of $\Lambda$ hypernuclei
in the RMF model. Most studies of hypernuclei in the RMF model
are performed by treating meson-hyperon couplings as
phenomenological parameters, which are determined by
experimental data~\cite{rmf2,rmf3,rmf4,rmf5}.
In the present calculation, the basic parameters are the quark-meson
couplings, and the effective meson-hyperon couplings are obtained
self-consistently at the quark level. The resulting $\Lambda$
single-particle energies with the parameter set B are very close
to the experimental values as shown in Fig.~\ref{fig:spe}.

\begin{figure}[htb]
\includegraphics[bb=30 90 520 520, width=8.6 cm, clip]{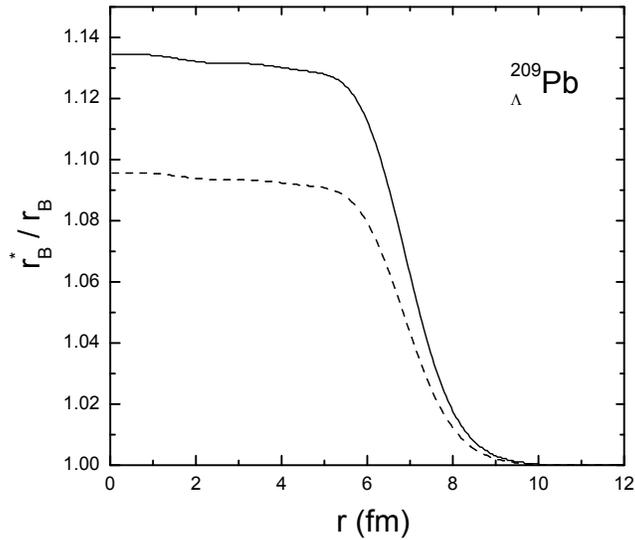}
\caption{The ratio of the baryon rms radius in $^{209}_{\Lambda}$Pb
to that in free space $r_B^{\ast}/r_B$ ($B=N,\,\Lambda$)
as a function of the radial coordinate $r$ with the parameter set B.
The results of $N$ and $\Lambda$ are shown by the solid
and dashed lines, respectively.}
\label{fig:ra}
\end{figure}

\begin{figure}[htb]
\includegraphics[bb=50 90 520 480, width=8.6 cm, clip]{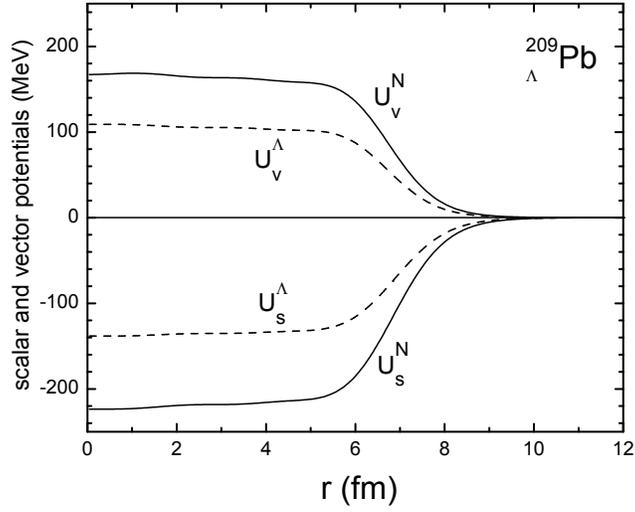}
\caption{The scalar and vector potentials
of the neutron (solid line) and $\Lambda$ (dashed line)
in $^{209}_\Lambda\rm{Pb}$
as functions of the radial coordinate $r$ with the parameter set B.}
\label{fig:uv}
\end{figure}

It is interesting to discuss the medium modifications of baryon properties
in $\Lambda$ hypernuclei. Using the local density approximation,
the baryon properties at the radial coordinate $r$ are obtained
through the value of $\sigma \left( r\right)$.
The dependence of baryon properties on the $\sigma$ mean field
are obtained at the quark level in the FL model.
In Fig.~\ref{fig:ra}, we show the ratio of the baryon rms radius
in $^{209}_{\Lambda}$Pb to that in free space as a function of the radial
coordinate $r$ with the parameter set B.
It is found that both $r_N^{\ast}$ and $r_{\Lambda}^{\ast}$
increase significantly at the center of $^{209}_{\Lambda}$Pb,
while they decrease to the values in free space at the surface.
In Fig.~\ref{fig:uv}, we plot the scalar and vector potentials
of the neutron and $\Lambda$ at $1s_{1/2}$ state
in $^{209}_\Lambda\rm{Pb}$. It is seen that both
$U_s^{\Lambda}$ and $U_v^{\Lambda}$ are smaller than
$U_s^N$ and $U_v^N$. At the center of $^{209}_{\Lambda}$Pb,
the attractive scalar potentials are mostly canceled by the
repulsive vector potentials, so that the difference between
the scalar and vector potentials is about $-56$ MeV for the neutron
and $-29$ MeV for the $\Lambda$ hyperon.

To examine the influence of $\Lambda$ on the nucleons in hypernuclei,
we list in Table~\ref{tab:spe} the calculated single-particle energies
of protons and neutrons in $^{41}_\Lambda\rm{Ca}$ with a $1s_{1/2}$ $\Lambda$
and compare them with those in $^{40}\rm{Ca}$.
It is shown that the existence of the $\Lambda$ hyperon does not
cause observed changes of the single-particle energies of protons and neutrons.
The appearance of the $\Lambda$ increases the baryon density
in the hypernucleus, so that both scalar and vector potentials are enhanced
in comparison with the nucleus without the $\Lambda$ hyperon.
However, the enhancements of the scalar and vector potentials can be
mostly canceled by each other, so that no significant effect is left
to the nucleon single-particle motion. This is consistent with
the results of the RMF models~\cite{rmf2,rmf3,rmf4,rmf5}.

It is also interesting to discuss the applicability of the present
model to $\Sigma$ and $\Xi$ hypernuclei. We note that both $\Lambda$
and $\Sigma^0$ are made up of three quarks ($uds$), which couple with
a scalar field to form the baryon in the FL model.
To obtain the mass difference between $\Lambda$ and $\Sigma$,
we take into account the spin-dependent interactions between quarks
by adding $E^B_{\text{spin}}$ in Eq.~(\ref{eq:nms1}).
Then the effective baryon mass is given by
$M_B^{\ast}\left(\sigma\right)=\sqrt{\left(E_B+E^B_{\text{spin}}\right)^{2}
 -\sum_{f}\langle \mathbf{p}_f^{2}\rangle }$,
where $E^B_{\text{spin}}$ is determined by the baryon mass in free space.
This treatment is very like the method used in the QMF model~\cite{qmf02},
and it is similar to adjusting the parameter $Z_Y$ by hyperon masses
in an earlier version of the QMC model~\cite{qmcl2}.
Using this simple method, we obtain the potential depth of a $\Sigma$
in saturated nuclear matter to be around $-28$ MeV with the parameter set B,
which is very close to the value of $\Lambda$ case.
Therefore, $\Sigma$ hypernuclei in the present model could be
bound by an amount similar to $\Lambda$ hypernuclei.
Such a result is not consistent with the repulsive $\Sigma$ potential
suggested by recent experimental studies.
Hopefully, the model can be improved by including the quark-quark hyperfine
interactions as done in the QMC model~\cite{qmc08}.
For $\Xi$ hypernuclei, we obtain the potential depth of a $\Xi$
in saturated nuclear matter to be around $-14$ MeV with the parameter set B,
and consequently $\Xi$ hypernuclei are weakly bound in the present model.
We note that the parameter set A yields qualitatively similar results for $\Sigma$
and $\Xi$ hypernuclei.

\section{Conclusion}
\label{sec:4}

In this paper, we have developed a model for the description of $\Lambda$ hypernuclei
and for investigating the medium modifications of baryon properties.
We have used the FL model to describe the nucleon and $\Lambda$ in nuclear matter
and hypernuclei. In the FL model, the baryon is described as a bound state of
three quarks in a nontopological soliton formed by a scalar composite gluon field.
The baryons interact through the self-consistent exchange of scalar and vector mesons
generated by the nuclear environment. The mesons, which are treated as classical
fields in the spirit of the QMC model, couple directly to the quarks inside the baryons.
The quark degrees of freedom are explicitly considered in the model; therefore,
it enables us to investigate the medium modifications of baryons in a nuclear many-body
system and provide a reasonable description of the system in a consistent manner.

We have presented the calculated results of $\Lambda$ hypernuclei and the medium
modifications of baryon properties with two sets of parameters, which correspond
to the two limiting cases in the parameter space of the FL model.
The parameters in the FL model are constrained by reproducing free baryon properties,
while the quark-meson couplings are determined by fitting saturation properties
of nuclear matter. Therefore, no more adjustable parameters exist when we calculate
the properties of $\Lambda$ hypernuclei. It has been found that
the resulting $\Lambda$ single-particle energies with the parameter set B are very
close to the experimental values, while those with set A are slightly underestimated.
We have found that the properties of the nucleon and $\Lambda$ are
significantly modified in nuclear medium. At normal nuclear matter density,
the nucleon radius increases by about $13\%$ ($30\%$) with the parameter set B (A),
while the $\Lambda$ radius increases by about $9\%$ ($23\%$).
In the $\Lambda$ hypernucleus, both nucleon and $\Lambda$ radii increase significantly
from the surface to the interior. It is gratifying to note that the present model
is able to provide a reasonable description of $\Lambda$ hypernuclei without
adjusting parameters, and at the same time it predicts a significant increase of
the baryon radius in nuclear medium.

It is important to point out the limitations of the present model
and possible directions for further improvements.
We have examined its applicability to $\Sigma$ and $\Xi$ hypernuclei.
Using a simple method for including spin-dependent interactions,
we found that $\Sigma$ hypernuclei are bound by an
amount similar to $\Lambda$ hypernuclei and $\Xi$ hypernuclei are
weakly bound in the present model. This is not consistent with
the repulsive $\Sigma$ potential suggested by recent experimental studies.
According to recent developments of the QMC model, the quark-quark
hyperfine interaction due to the gluon exchange plays an
important role in explaining why $\Lambda$ hypernuclei are
bound whereas $\Sigma$ hypernuclei are not.
In further work, we plan to include the quark-quark hyperfine interaction
in the FL model and explore its medium modification.
It is very interesting to investigate and compare the influence of the
hyperfine interaction on properties of in-medium baryons and hypernuclei.


\section*{Acknowledgments}

This work was supported in part by the National Natural Science Foundation
of China (Grants No. 11075082 and No. 11375089).


\newpage

\begin{table}[tbp]
\caption{The quark-meson couplings ($q=u,d$) and the nuclear matter properties
corresponding to the parameter sets A and B.
The saturation density and the energy per
particle are denoted by $\protect\rho_0$ and $E/A$, the symmetry energy
by $a_{\mathrm{sym}}$, the incompressibility by $K$, and the effective
nucleon mass by $M_N^{\ast}$.}
\label{tab:matter}
\vspace{0.0cm}
\begin{center}
\begin{tabular}{ccccccccc}
\hline\hline
& $ g_{\sigma}^q $ & $g_{\omega}^q$ & $g_{\rho}^q$
& $\rho_0$ & $E/A $ & $a_{\mathrm{sym}}$ & $K$ & $M_N^{\ast }/M_N$ \vspace{-0.2cm} \\
& & & & (fm$^{-3})$ & (MeV) & (MeV) & (MeV) &  \\ \hline
Set A & 17.023& 3.629& 3.96& 0.15 & -16.0 & 32 & 317 & 0.69 \\
Set B & 9.275&3.056 &4.15 & 0.15 & -16.0 & 32& 338& 0.77 \\ \hline\hline
\end{tabular}%
\end{center}
\end{table}

\begin{table}[tbp]
\caption{The calculated single-particle energies of protons and neutrons
in $^{41}_\Lambda\rm{Ca}$ and $^{40}\rm{Ca}$.
The $\Lambda$ hyperon in $^{41}_\Lambda\rm{Ca}$ is at $1s_{1/2}$ state.
The experimental data are taken from Ref.~\protect\cite{EXPSO}.
All energies are in MeV.}
\label{tab:spe}\vspace{0.0cm}
\begin{center}
\begin{tabular}{cccccccccccc}
\hline\hline
&& \multicolumn{2}{c}{Set A} &  & \multicolumn{2}{c}{Set B} &  & \multicolumn{2}{c}{Expt.} \\
\cline{3-4}\cline{6-7}
&& $^{41}_{\Lambda}\text{Ca}$ & $^{40}\text{Ca}$ &
 & $^{41}_{\Lambda}\text{Ca}$ & $^{40}\text{Ca}$ &  & $^{40}\text{Ca}$ \\ \hline
Proton
&$1s_{1/2}$ & 38.2 & 38.5 &  & 36.3 & 36.5 &   & 50$\pm$11 &
\\
&$1p_{3/2}$ & 25.8 & 25.9 &  & 25.1 & 25.1 &   & 34$\pm$6 &
 \\
&$1p_{1/2}$ & 23.4 & 23.5 &  & 23.5 & 23.6 &   & 34$\pm$6 &
\\
&$1d_{5/2}$ & 13.0 & 12.9 &  & 12.8 & 12.6 &   & 15.5
\\
&$1d_{3/2}$ &  9.1 &  9.0 &  & 10.1 & 10.0 &   & 8.3 \\
&$2s_{1/2}$ &  8.2 &  8.2 &  &  7.8 &  7.8 &   & 10.9  \\
Neutron
&$1s_{1/2}$ & 46.3 & 46.7 &  & 44.5 & 44.7 &   &
50.0 \\
&$1p_{3/2}$ & 33.6 & 33.7 &  & 32.9 & 33.0 &   &
30.0 \\
&$1p_{1/2}$ & 31.2 & 31.3 &  & 31.4 & 31.4 &   &
27.0 \\
&$1d_{5/2}$ & 20.4 & 20.3 &  & 20.3 & 20.2 &   & 21.9
\\
&$1d_{3/2}$ & 16.5 & 16.4 &  & 17.6 & 17.5 &   & 15.6 \\
&$2s_{1/2}$ & 15.5 & 15.6 &  & 15.3 & 15.3 &   & 18.2 \\
\hline\hline
\end{tabular}%
\end{center}
\end{table}


\end{document}